\documentclass[twocolumn,aps,floatfix,superscriptaddress,prl,nobibnotes,longbibliography]{revtex4-1}

\usepackage{graphicx}
\usepackage{amsmath,amsthm,amssymb,amsfonts}

\usepackage[USenglish]{babel}
\usepackage[OT4]{fontenc}
\usepackage[bbgreekl]{mathbbol}
\usepackage{bm}
\usepackage{color}
\usepackage{lipsum}
\usepackage[utf8]{inputenc}

\usepackage[colorlinks=true,linkcolor=blue,citecolor=blue]{hyperref}
\usepackage{accents}
\newlength{\dhatheight}

\def\e{\mathrm{e}}

\def\rmi{\mathrm{i}}

\def\H{\mathcal{H}}

\newcommand{\mbf}[1]{%
    \mathbf{#1}}

\begin{document}

\title{Bulk topological proximity effect in multilayer systems}

\author{Jaromir Panas}
\email{Jaromir.Panas@fuw.edu.pl}
\affiliation{Institut f\"ur Theoretische Physik, Goethe-Universit\"at, 60438 Frankfurt am Main, Germany}

\author{Bernhard Irsigler}
\affiliation{Institut f\"ur Theoretische Physik, Goethe-Universit\"at, 60438 Frankfurt am Main, Germany}

\author{Jun-Hui Zheng}
\affiliation{Institut f\"ur Theoretische Physik, Goethe-Universit\"at, 60438 Frankfurt am Main, Germany}
\affiliation{Center for Quantum Spintronics, Department of Physics, Norwegian University of Science and Technology, NO-7491 Trondheim, Norway}

\author{Walter Hofstetter}
\affiliation{Institut f\"ur Theoretische Physik, Goethe-Universit\"at, 60438 Frankfurt am Main, Germany}

\date{\today}

\begin{abstract}
We investigate the bulk topological proximity effect in multilayer hexagonal lattice systems by which one can introduce topological properties into a system composed of multiple trivial layers by tunnel coupling to a single nontrivial layer described by the Haldane model.
This phenomenon depends not only on the number of layers but also on their arrangement, which can lead to the emergence of {\it dark states} in multilayer systems.
The response of a trivial system to the proximity of a topological insulator appears to be highly nonlocal, in contrast to the proximity effect observed in context of superconductivity.
Furthermore, for a wide range of parameters our system is semimetallic with multiple Dirac points emerging in the Brillouin zone.
\end{abstract}

\pacs{}

\maketitle

The proximity effect of superconductivity is a well studied phenomenon.
By bringing a superconducting material with a finite, local, $U(1)$ order parameter into contact with a nonsuperconducting one, the order parameter is inherited into the bulk of the nonsuperconducting material.
This effect has an analogue for topological materials~\cite{Hasan2010}, even though they do not possess a local order parameter.
Hsieh et al.~\cite{Hsieh2016} describe the effect of a Chern insulator with Chern number $C=1$ coupled to a topologically trivial material.
Due to coupling the trivial material becomes topologically nontrivial with an opposite Chern number $C=-1$.
It is important to emphasize that this is bulk physics and must be differentiated from topological edge states.
The effect has been studied for bilayer hexagonal lattice systems~\cite{Haldane1988}  in Refs.~\cite{Zheng2018,Cheng2019}.
Reference~\cite{Zheng2018} introduces a topological invariant for open systems which makes it possible to compute Chern number of a subsystem, e.g., for single layer.
This technique gives evidence of the emergence of the $C=-1$ Chern number in the trivial layer.
Another study investigates bilayers of two Haldane insulators with opposite Chern numbers and found various topological many-body phases, especially if two-body interactions in one layer are applied~\cite{Sorn2018}.
A spinful bilayer system of stacked Kane-Mele layers~\cite{Kane2005} has been investigated in Ref.~\cite{Mishra2019}, and various types of bulk proximity effects involving topologically nontrivial systems coupled to few topologically trivial layers have been recently studied in real materials, both theoretically~\cite{Zeng2019,Zollner2019} as well as experimentally~\cite{Shoman2015}.

The main difference between the superconducting and the bulk topological proximity effects is that the latter does not possess a local order parameter.
Moreover, the proximity of the nontrivial layer induces topological properties with opposite chirality~\cite{Hsieh2016,Cheng2019}, as manifested by the sign of the topological invariant~\cite{Zheng2018}.
The full system is then topologically trivial.
Following this it is a priori not evident how, e.g., two equal trivial layers would 'compensate' the Chern number of a third, nontrivial one.
By studying multilayer systems we aim to get a better understanding of the bulk topological proximity effect and how it differs from the proximity effect of superconductivity.

\textit{Three-layer system --} We begin our investigation with a detailed study of a three-layer system, which is described by a tight-binding model with spinless, noninteracting fermions on a honeycomb lattice.
In our investigation we allow for the following terms in the Hamiltonian: (i) the nearest-neighbor (NN) hopping with an amplitude $t_1$, (ii) the next-nearest-neighbor (NNN) hopping with an amplitude $t_2$ and an associated change of phase $\Phi$, (iii) the staggered potential with an amplitude $m$, and (iv) the interlayer hopping with an amplitude $r$.
The general form of the Hamiltonian in momentum space for such a system has the form
\begin{equation}\label{eqn:main:Ham}
\H(\mathbf{k})=\begin{pmatrix}
\vec{d}_1(\mathbf{k}) \cdot \vec{\sigma} & r \sigma_0 & \mathbb{0} \\
r  \sigma_0 & \vec{d}_2(\mathbf{k}) \cdot \vec{\sigma} & r  \sigma_0 \\
\mathbb{0} & r  \sigma_0 & \vec{d}_3(\mathbf{k}) \cdot \vec{\sigma} \\
\end{pmatrix}.
\end{equation}
Here,  $\mbf{k}=(k_x,k_y)$ is a quasimomentum in the two-dimensional Brillouin zone (BZ), $\sigma_0$ and $\mathbb{0}$ are the 2$\times$2 unit and zero matrix, respectively, and $\vec{\sigma} = (\sigma_0,\sigma_x,\sigma_y,\sigma_z)$ is a four-component vector of unit and Pauli matrices.
Each $2\times 2$ diagonal block represents a single layer.
The properties of a given layer are captured by the four-dimensional vector $\vec{d}_i(\mbf{k})$, which represents the Hamiltonian of the decoupled layer $i$ in the Bloch sphere representation and which is given by~\cite{Cheng2019}
\begin{equation}\label{eqn:main:d}
\vec{d}_i(\mathbf{k})
=
\begin{pmatrix}
-2 t_{2;i} \cos(\Phi_i) \sum_{i=1}^3 \cos(\mbf{k}\cdot\mbf{b}_i)\\
-t_{1;i} \sum_{i=1}^3 \cos(\mbf{k}\cdot\mbf{a}_i) \\
-t_{1;i} \sum_{i=1}^3 \sin(\mbf{k}\cdot\mbf{a}_i) \\
m_i-2 t_{2;i} \sin(\Phi_i) \sum_{i=1}^3 \sin(\mbf{k}\cdot\mbf{b}_i)
\end{pmatrix}.
\end{equation}
The vectors $\mbf{a}_i$ and $\mbf{b}_i$ ($|\mbf{a}_i|=1$) link the NNs and NNNs within the honeycomb lattice, respectively.

In the following we will restrict our considerations to the case where one of the layers is described by the Haldane model~\cite{Haldane1988} with $\Phi=\pi/2$ and the remaining two layers are graphene layers in the tight-binding model with staggered potential.
Therefore, the possible values of $\vec{d}_i(\mathbf{k})$ are limited to $\vec{d}_i(\mathbf{k})\in \{ \vec{d}_g(\mathbf{k}),\vec{d}_h(\mathbf{k})\}$ where $t_{1;g}=4$, $t_{2;g}=0$, $m_g=1$ and $t_{1;h}=4$, $t_{2;h}=1$, $\Phi_h=\pi/2$, $m_h=0$ (we use $t_{2;h}$ as a unit of energy).
The values have been chosen based on Ref.~\cite{Zheng2018} for comparison but the conclusions are general as discussed below.

\begin{figure}
\resizebox{1.0\columnwidth}{!}{
\includegraphics[scale=1]{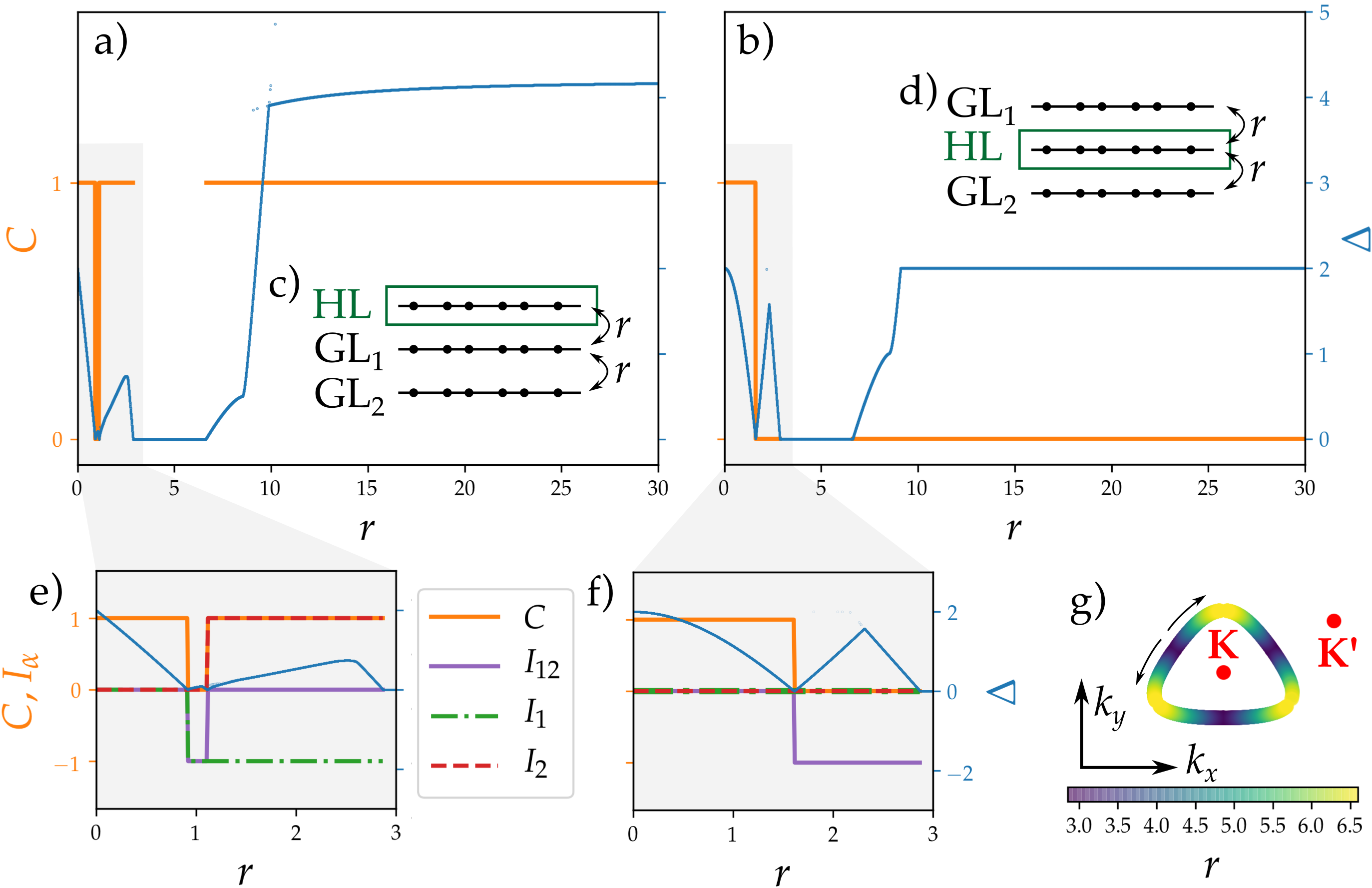}}
\caption{Band gap (blue) and Chern number (orange) of the full three-layer system for a) configuration HGG  and b) configuration GHG, which are schematically represented in c) and d), respectively. In e) and f) the topological indices $I_1$ and $I_2$ for the individual GLs as well as $I_{12}$ for subsystem consisting of both GLs are plotted for smaller range of coupling strengths. For both configurations at intermediate couplings $2.9 \lesssim r \lesssim 6.6$ the gap is closed at three pairs of Dirac points that lie on a closed contour in the BZ, which is represented in g). With increasing $r$ the Dirac points move along the contour as depicted by arrows. Used parameters are $t_1=4$ in all layers, $m=1$ in GLs, and $t_2=1$ with $\Phi=\pi/2$ in the HL.}
\label{fig:main:bulk}
\end{figure}

We consider two arrangements of the three layers: Configuration HGG, with the Haldane layer (HL) being on top of the two graphene layers (GLs), as depicted in Fig.~\ref{fig:main:bulk} c), where $\vec{d}_1 = \vec{d}_h$ and  $\vec{d}_2 = \vec{d}_3 = \vec{d}_g$, or configuration GHG, with the HL sandwiched between two GLs as shown in Fig.~\ref{fig:main:bulk} d), where $\vec{d}_2 = \vec{d}_h$ and  $\vec{d}_1 = \vec{d}_3 = \vec{d}_g$.
We assume that the layers are parallel to the $xy$~plane, AA-stacked~\cite{Charlier1991} in the $z$~direction, and that the system is half-filled.

To determine topological properties of the system we calculate its Chern number $C$
\begin{equation}\label{eqn:main:C}
C=\sum_n \int_{BZ} \mathrm{d}^2 k\frac{\epsilon^{\mu\nu}}{2\pi\mathrm{i}} \left\langle  \partial_{k_\mu} \psi_n(\mathbf{k}) | \partial_{k_\nu} \psi_n (\mathbf{k}) \right\rangle.
\end{equation}
Here $\epsilon^{\mu\nu}$ is an antisymmetric tensor, $\psi_n$ are the occupied (momentum-) eigenstates of the system, and $\mu,\nu=x,y$.
We use Fukui's method~\cite{Fukui2005} on the discretized BZ to numerically obtain $C$.
To determine topological properties of a subsystem $\alpha$ we employ the topological invariant $I_\alpha$ developed in Ref.~\cite{Zheng2018}.
In this method eigenstates $\psi_n$ in \eqref{eqn:main:C} are replaced by the eigenstates with the highest eigenvalues of the single-particle density matrix of the given subsystem $\alpha$.
Here, we consider the subsystem to either consist of a single GL or of the two GLs together, \textit{excluding} the HL. 
In Fig.~\ref{fig:main:bulk}~e) $I_1$ and $I_2$ are topological indices of the two GLs $\mathrm{GL}_1$ and $\mathrm{GL}_2$, respectively, while $I_{12}$ is the topological index of a subsystem consisting of both $\mathrm{GL}_1$ and $\mathrm{GL}_2$ taken together.
The results of numerical calculations are presented in Fig.~\ref{fig:main:bulk}~a) and e) for configuration HGG and in Fig.~\ref{fig:main:bulk}~b) and f) for configuration GHG.
For vanishing interlayer tunneling $r=0$ in both configurations, the gap is open due to the staggered potential $m$ and the NNN hopping $t_2$, and the full system is topologically nontrivial.
Correspondingly, the Chern number is $C=1$, which is a sum of topological indices of each layer $C=I_{HL}+I_1+I_2$, with $I_{HL}=1$, $I_1=0$, and $I_2=0$.
As we increase the coupling strength, in both configurations, the gap decreases and eventually closes at the $\mbf{K}=\frac{2\pi}{3}\left(1,\frac{1}{\sqrt{3}}\right)$ point in the BZ at critical value $r_{1}$, which marks a phase transition of the system to a topologically trivial state with $C=0$.
Note that the values of $r_{1}$ are different for the two configurations.
We also note that the behavior for $r>r_{1}$ is significantly different in the two configurations.

In configuration HGG, the system remains in the topologically trivial state for a small range of coupling strengths~$r$.
At critical value $r_{2}$, the gap closes at the $\mbf{K'}=\frac{2\pi}{3}\left(1,-\frac{1}{\sqrt{3}}\right)$ point in the BZ leading to another phase transition.
We also note that the two GLs change the values of their respective topological indices $I_\alpha$ sequentially.
For weak coupling, the indices have values $I_1=I_2=0$.
Then at $r_{1}$, the $I_1$ index of the layer neighboring the HL changes to $I_1=-1$ and after the second phase transition at $r_{2}$ the $I_2$ index of the last GL changes to $1$.
The topological index $I_{12}$ of the two GLs follows $I_{12}=I_1+I_2$.
The HL topological index $I_\mathrm{HL}=1$ remains unchanged.
Therefore, we have $C=I_\mathrm{HL}+I_{12}=I_\mathrm{HL}+I_{1}+I_2$.

In configuration GHG, the system remains in a topologically trivial state for all $r>r_{1}$.
The individual behavior of the indices $I_\alpha$ is also different.
The initial values $I_1=I_2=0$ and $I_\mathrm{HL}=1$ remain the same for all $r$ where they are properly defined.
However, the index of two GLs together $I_{12}$ does change to $I_{12}=-1$ at $r_{1}$.
Therefore, we have $C=I_\mathrm{HL}+I_{12} \neq I_\mathrm{HL}+I_{1}+I_2$.

The difference between the two configurations HGG and GHG is explained by the emergence of a \textit{dark state} in the sandwiched configuration GHG.
We choose this naming in correspondence with dark states in open quantum systems \cite{Breuer2002,Griessner2006,Kraus2008}, it is also used for a closely related effect~\cite{Lacki2020}.
Dark states in our system can be engineered knowing the eigenstates ${v}_\pm(\mbf{k})$ of the graphene Hamiltonian $\vec{d}_g(\mbf{k})\cdot\vec{{\sigma}}$.
We notice, that the six-component statevector $[{v}_\pm^\dag(\mbf{k}), 0,0, -{v}_\pm^\dag(\mbf{k})]^\dag$ is an eigenvector of the Hamiltonian~\eqref{eqn:main:Ham} in the configuration GHG with the same energy as the eigenstate ${v}_\pm(\mbf{k})$ of graphene.
We therefore obtain a state that is completely decoupled from the HL, due to the vanishing amplitude at the central layer, and insensitive to the coupling strength $r$.
On the other hand, states $[{v}_\pm^\dag(\mbf{k}), 0,0, {v}_\pm^\dag(\mbf{k})]^\dag$ are coupled stronger to the HL, compared to a single-layer state, with an effective coupling strength $r^\mathrm{eff}=\sqrt{2}r$.
As a result, the configuration GHG can be mapped onto a bilayer problem, which was investigated in Ref.~\cite{Zheng2018,Cheng2019}, and a decoupled, effective GL.
The above arguments no longer hold for the configuration HGG where such an eigenstate with vanishing amplitude at the HL does not exist.
We note that the dark states occur due to the layer stacking symmetry, and is not inherently related to topological phenomena.
Nevertheless, their presence significantly affects the global topological properties of the system.

In both considered configurations at intermediate values of $r$ the gap closes again at $r_< \approx 2.9$, remains closed for a certain range of coupling strengths, and finally reopens at $r_>\approx 6.6$ and remains open for $r\rightarrow\infty$.
In contrast to previous gap closing instances, this one does not occur at the $\mbf{K}$ or $\mbf{K'}$ points but rather at points from a subset $\mathcal{D}$ of the BZ given by
\begin{equation}\label{eqn:main:loop}
\mathcal{D}=\left\{\mbf{k} : \vec{d}_h(\mbf{k})=\vec{d}_g(\mbf{k})\right\}.
\end{equation}
This set forms a closed contour in the BZ depicted in Fig.~\ref{fig:main:bulk} g).
At $r_<$ the gap closes at points $\mbf{k}\in\mathcal{D}$ that lie on the lines connecting the $\mbf{K}$ point and its three neighboring $\mbf{K'}$ points.
The system becomes semimetallic with three pairs of Dirac points, one for each $\mbf{K'}$ neighbor of $\mbf{K}$.
The semimetallic properties emerge even though our system is exposed to both staggered potential and gauge field. In graphene, either of these open a gap.
In each pair the Dirac points have opposite vorticity.
As $r$ is increased the pair of Dirac points that were initially created at the same $\mbf{k}\in\mathcal{D}$  move away from one another along the $\mathcal{D}$ contour, as depicted in Fig.~\ref{fig:main:bulk} g) with arrows.
Eventually, at $r_>$, the Dirac points of opposite vorticity annihilate in new pairs, and the gap opens again.
A more detailed explanation of this mechanism can be found in the Supplemental Material and also in Ref.~\cite{Sorn2018}.

We have also numerically checked that the features described above for the three-layer systems remain robust against small modifications of parameters, even when we vary NN hopping amplitude $t_1$, NNN hopping amplitude $t_2$, phase shift $\Phi$, staggered potential $m$, and interlayer coupling $r$ for each layer independently.
We have also tested robustness of the Dirac points at intermediate values of $r$ against the next-next-nearest-neighbor hopping.
Contrary to the suggestion of Ref.~\cite{Sorn2018} we find that it only shifts the Dirac points from the contour $\mathcal{D}$, without opening the gap (see the Supplemental Material).

To better understand the observations for different configurations of layers we investigate the system in the strong coupling limit.
We perform a perturbation expansion with respect to terms $m, t_2 \ll r$, details of which can be found in the Supplemental Material.
When the coupling is sufficiently strong, the spectrum splits into pairs of bands separated by an energy offset of order $r$.
The dispersion of different bands reads
\begin{equation}\label{eqn:main:ene}
E_\pm(\mbf{k},\kappa_z) = - 2 r \cos(\kappa_z) + \epsilon_\pm (\mbf{k},\kappa_z),
\end{equation}
where $\kappa_z\in\{\frac{\pi}{4},\frac{2\pi}{4},\frac{3\pi}{4}\}$ and $\epsilon_\pm(\mbf{k},\kappa_z)$ are the eigenenergies of the Haldane models with renormalized parameters $t_1^\mathrm{eff}(\kappa_z)=t_1$, $t_2^\mathrm{eff}(\kappa_z)$, $m^\mathrm{eff}(\kappa_z)$, $\Phi^\mathrm{eff}(\kappa_z)=\Phi=\pi/2$.
Different pairs of bands vary with respect to the $z$ dependence of their wavefunction.
Because of this effective Haldane model description, each band can have a finite Chern number.
Since the sum of the Chern numbers of bands from a given pair is zero, the topological properties of the entire half-filled system will be determined by the pair that lies closest to the Fermi energy $E_F$.
If $E_F$ lies between the two bands of such a pair, the system as a whole will acquire a finite Chern number $C$ of the lower band.
At half-filling in strong coupling limit only the states with $\kappa_z=\pi/2$ (with energies $E_\pm(\mbf{k},\kappa_z)=\epsilon_\pm(\mbf{k},\kappa_z)$) are relevant for the Chern number of the full system.

An important observation is that the wavefunctions are delocalized with respect to the layer index.
This poses an issue for the interpretation of the topological index $I_\alpha$ and leads to $I_1+I_2\neq I_{12}$ in the GHG configuration.
Trivial or nontrivial topology is a property of a band rather than a layer.
Therefore, unless bands are approximately localized on specific layers, using the topological index $I_\alpha$ leads to inconsistencies.
While in Ref.~\cite{Zheng2018} bands could be approximately associated with layers thanks to the weak coupling and different energy scales in the HL and the GL, this is no longer the case in our three-layer system.
In the configuration GHG, even in the weak coupling limit, the bands are always delocalized between the two GLs, due to their degeneracy at $r=0$ and the symmetry of the layer arrangement.
As a result, bands can be associated either with a Haldane layer or with the subsystem composed of two graphene layers, resulting in values of $C=I_\text{HL}+I_{12}$, $I_\text{HL}=1$, and $I_{12}=-1$, but not with any of the graphene layers separately, giving $I_1=I_2=0$ and $I_{12}\neq I_1+I_2$, c.f.~Fig.~\ref{fig:main:bulk}.
We note that at $r=0$ the GLs are degenerate with each other but not with the HL.
It is subject of future studies to determine a quantitative method for deciding whether a band can be associated with a subsystem in a manner sufficient for $I_{\alpha}$ to give meaningful results.

Finally, we comment on the dependence of our observations on the values of the parameters.
As mentioned above, small changes, even such that reduce the symmetry of the layer arrangement, do not affect the observed behavior qualitatively.
Upon larger changes, the observed physics will change.
For example, changing $t_1$ affects the critical values $r_<$ and $r_>$, and the onset of semimetallic properties can occur before the first topological phase transition.
Another example is when changing $t_2$ with respect to $m$.
In a regime, where $m^\text{eff}/t_2^\text{eff}>3\sqrt{3}$, the system is topologically trivial even in the strong coupling limit of the HGG configuration.
Nevertheless, the underlying mechanisms remain similar.


\textit{Multilayer system --} We generalize our study of a three-layer system to the one with $L$ layers, $L-1$ of which are trivial.
Increasing the coupling from zero we expect a series of gap closings and reopenings occuring interchangeably at the $\mathbf{K}$ and $\mathbf{K}'$ special points in the Brillouin zone. 
Each leads to a band inversion and a change of the topological state of the half-filled system between ones with $\mathcal{C}=1$ and $\mathcal{C}=0$.
Due to the simple form of the Hamiltonian the critical coupling strengths can be easily calculated numerically.
These gap closings are independent of the ones responsible for the onset of semimetallic properties, which occur on the contour $\mathcal{D}$ defined in \eqref{eqn:main:d}.
Due to the larger number of layers, the range of coupling strengths where the system is semimetallic and the number of emerging Dirac points appearing on $\mathcal{D}$ will be larger.
However, unlike the three-layer system, a multilayer system can become metallic for certain layer arrangements.
This will occur when multiple dark states emerge that are decoupled from the HL but coupled to one another.

\begin{figure}
\resizebox{1.0\columnwidth}{!}{
\includegraphics[scale=1]{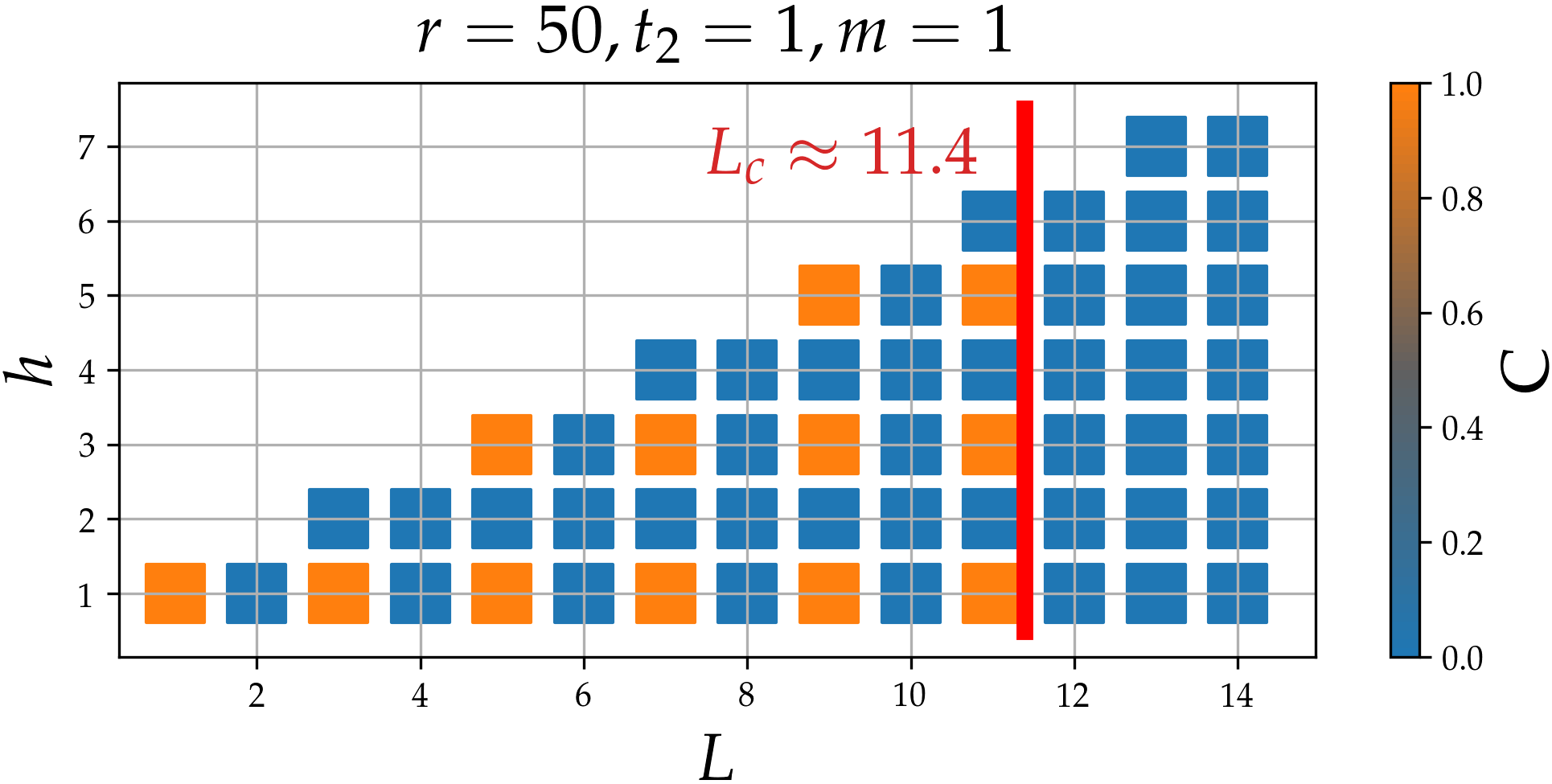}}
\caption{Chern number $C$ as a function of number of layers $L$ and the layer index of the HL $h$ in the strong coupling limit. All cases with even $L$ and $h$ are topologically trivial as explained in the text. For $L>L_c\approx 11.4$ the system is also topologically trivial due to effective ratio $m^\mathrm{eff}/t_2^\mathrm{eff}> 3\sqrt{3}$.}
\label{fig:main:multi}
\end{figure}

To get a better intuition for the multilayer case it is insightful to study the model in the strong coupling limit $r/L\gg t_1, t_2, m$.
Similarly as in the three-layer case we obtain an effective Hamiltonian with $L$ pairs of decoupled bands.
For detailed derivations see the Supplemental Material.
The dispersion  relation is given in Eq.~\eqref{eqn:main:ene} but with $\kappa_z\in \{ \frac{\pi}{L+1},\ldots,\frac{L\pi}{L+1}\}$ and with $\epsilon_\pm (\mathbf{k},\kappa_z)$ being bands of an effective Haldane model with $t_1^{\mathrm{eff}}(\kappa_z)=t_1$, $t_2^{\mathrm{eff}}(\kappa_z)=|\mathcal{N}|^2\sin^2(\kappa_z h) t_2$, $\Phi^{\mathrm{eff}}(\kappa_z)=\Phi=\pi/2$, and $m^{\mathrm{eff}}= \left( 1- |\mathcal{N}|^2 \sin^2(\pi h/2) \right) m$.
$h$ is the index of the HL and $\mathcal{N}$ is the normalization factor, which for $\kappa_z=\pi/2$ and odd $L$ reads $|\mathcal{N}|^2={2}/({L+1})$.
The Chern number of the system, shown in Fig.~\ref{fig:main:multi} for $r=50$, is determined by the bands below the Fermi energy.
We can make three observations.
(i) In the half-filled case with $E_F=0$ systems with an even number of layers $L$ will be topologically trivial, due to even number of occupied bands.
(ii) For an odd number of layers $L$ there exist two bands with $\kappa_z=\pi/2$, which determine the topological properties of the half-filled system.
In particular, the system will be topologically nontrivial if the effective parameters satisfy $m^\mathrm{eff}(\pi/2)/t_2^\mathrm{eff}(\pi/2) <3\sqrt{3}$~\cite{Haldane1988}, which translates to a critical value $L_c=6\sqrt{3}+1 \approx 11.4$.
(iii) All bands with $\kappa_z h = n \pi$ where $n\in \mathbb{Z}$ correspond to the dark states due to the vanishing of $t_2^{\mathrm{eff}}(\pi/2)$ -- the HL has no influence on their dispersion.
We note that the Hilbert space can be split into a subspace of dark states and a subspace of states coupled to the HL, resulting in effectively two decoupled subsystems with different properties.

\textit{Conclusion --}
In conclusion, on an example of AA-stacked honeycomb lattice multilayer systems with a single topological layer we have presented important insights into the bulk topological proximity effect.
As previously discussed for the bilayer systems~\cite{Hsieh2016,Zheng2018,Cheng2019}, a single topological layer can induce topological response in the neighboring topologically trivial layers.
However, the bulk topological proximity effect stays in stark contrast to the superconducting proximity effect.
Most importantly it is highly nonlocal, as topological properties are features of bands rather than layers.
This leads to issues when trying to identify layer- or subsystem-specific topological indices.
Moreover the behavior of the system depends strongly on the layer arrangement owing to the emergence of dark states.
Finally, we observe that for a finite range of parameters the system becomes semimetallic.
We expect that our results can be easily generalized to other models, e.g., generalized Hofstadter and Kane-Mele models.
Studying interactions in these systems will be of interest in the near future and is promising from the perspective of state engineereing~\cite{Hsieh2017}.
We also expect that the systems used in our letter could be realized experimentally using a combination of shaken optical lattices~\cite{Struck2012,Sacha2012}, synthetic dimensions~\cite{Cheng2019} and spin-dependent optical potentials~\cite{Mandel2003}.

\begin{acknowledgments}
The authors acknowledge useful discussions with Monika Aidelsburger and Fabian Grusdt.
This work was supported by the Deutsche Forschungsgemeinschaft (DFG, German Research Foundation) under Project No. 277974659 via Research Unit FOR 2414. This work was also supported by the DFG via the high performance computing center LOEWE-CSC. The work was finalized while one of the authors (WH) was visiting the Institute for Mathematical Sciences, National University of Singapore in 2020. The visit was supported by the Institute.
\end{acknowledgments}


\newpage
\onecolumngrid
\appendix

\section{Strong coupling limit}
\label{app:SCL}

We consider a multilayer generalization of the three-layer system in tight-binding approximation considered in the main text.
There are $L$ honeycomb lattice layers AA-stacked along the $z$ axis each being parallel to the $xy$-plane and infinite.
The first $L_1$ and the last $L_2=L - L_1-1$ are GLs while the layer with index $h=L_1+1$ is a HL.
The full model in $\mbf{k}$-space is captured by the following Hamiltonian
\begin{equation}\label{eqn:app:1}
\H(\mathbf{k})=\begin{pmatrix}
\ddots & & & & \\
 & \vec{d}_g(\mathbf{k}) \cdot \vec{{\sigma}} & r\cdot \sigma_0 & \mathbb{0} & \\
 & r\cdot \sigma_0 & \vec{d}_h(\mathbf{k}) \cdot \vec{{\sigma}} & r\cdot \sigma_0 & \\
 & \mathbb{0} & r\cdot \sigma_0 & \vec{d}_g(\mathbf{k}) \cdot \vec{{\sigma}} & \\
 & & & & \ddots \\
\end{pmatrix},
\end{equation}
with terms defined in the main text.
As a sidenote we point to the fact that this Hamiltonian, while producing correct dispersions and Berry curvatures, is not periodic with respect to a shift to a neighboring copy of the BZ which would require additional unitary transformation~\cite{Zheng2018_2}.
We use this version of the Hamiltonian, as it has more concise notation.

We take the strong coupling limit of our system $r/L \gg t_1, t_2, m$, in which we can consider special properties of the Haldane layer as a small perturbation.
The unperturbed Hamiltonian $\H_0(\mbf{k})$ is then of form \eqref{eqn:app:1} with block terms $\vec{d}_g(\mathbf{k}) \cdot \vec{{\sigma}}$ on the diagonal, including the $h$ position.
The perturbation Hamiltonian $\H_I(\mbf{k})$ vanishes everywhere except for the $h$ diagonal block and is given by $(\vec{d}_h(\mathbf{k}) - \vec{d}_g(\mathbf{k}))  \cdot \vec{{\sigma}}$.
We first diagonalize the unperturbed Hamiltonian, which corresponds to solving a one-dimensional tight-binding chain with open boundary conditions (OBC) combined with diagonalizing $2\times 2$ matrix of graphene in $\mbf{k}$-space.
One can show that the eigenstates are product states of standing waves in $z$ direction with eigenstates of graphene in $xy$-plane
\begin{equation}\label{eqn:app:2}
v_\pm(\mathbf{k},\kappa_z)=\mathcal{N}[\alpha_\pm(\mbf{k}) \sin(\kappa_z),\beta_\pm(\mbf{k}) \sin(\kappa_z), \alpha_\pm(\mbf{k}) \sin(2\kappa_z), \beta_\pm(\mbf{k}) \sin(2 \kappa_z), \ldots \alpha_\pm(\mbf{k}) \sin(L \kappa_z),\beta_\pm(\mbf{k}) \sin(L \kappa_z)]^T,
\end{equation}
where $\mathcal{N}$ is the normalization factor and $\kappa_z \in \{ \frac{\pi}{L+1}, \frac{2 \pi}{L+1},\ldots, \frac{L \pi}{L+1}\}$.
The eigenvalues read
\begin{equation}\label{eqn:app:3}
E^{(0)}_\pm(\mbf{k} ,\kappa_z) = -2 r \cos (\kappa_z) + \epsilon_{g,\pm}(\mbf{k}),
\end{equation}
with $\epsilon_{g,\pm} (\mbf{k})$ being the energy bands of graphene.
The factors $\alpha_\pm(\mbf{k})$ and $\beta_\pm(\mbf{k})$ are components of the normalized eigenvectors of graphene.
Now we consider the correction due to $\H_I(\mbf{k})$. We focus on the first order correction to the eigenstates~\cite{Landau1977}
\begin{equation}\label{eqn:app:4} 
v^{(1)}_\pm(\mbf{k},\kappa_z) = \sum_{\kappa'_z\neq \kappa_z}\sum_{\gamma\in\{+,-\}} \left[ \frac{(v_\gamma(\mbf{k},\kappa'_z))^\dag \H_I(\mbf{k}) v_\pm(\mbf{k},\kappa_z)}{E_{\pm}(\mbf{k},\kappa_z)-E_\gamma(\mbf{k},\kappa'_z)} v_\gamma(\mbf{k},\kappa'_z) \right] + \frac{(v_\mp(\mbf{k},\kappa_z))^\dag \H_I(\mbf{k}) v_\pm(\mbf{k},\kappa_z)}{E_{\pm}(\mbf{k},\kappa_z)-E_\mp(\mbf{k},\kappa_z)} v_\mp(\mbf{k},\kappa_z).
\end{equation}
We notice that for sufficiently large $r/L \gg t_1, t_2, m, \ldots$ we obtain $|E_{\pm}(\mbf{k},\kappa_z)-E_\gamma(\mbf{k},\kappa'_z)| \gg |E_{\pm}(\mbf{k},\kappa_z)-E_\mp(\mbf{k},\kappa_z)|$ and only the terms with the same $\kappa_z$ will contribute significantly to $v^{(1)}_\pm(\mathbf{k},\kappa_z)$.
For higher order terms in perturbation expansion~\cite{Landau1977} this also holds. 
Therefore, we can neglect mixing between levels with different $\kappa_z$.
This simplifies the problem to a set of $L$ independent pairs of states, each defined by a different value of $\kappa_z$.
The effective Hamiltonian for such a pair reads
\begin{equation}\label{eqn:app:5}
\H^\mathrm{eff}(\mbf{k},\kappa_z) = -2 r \cos(\kappa_z) \mathbb{1} + \vec{d}_g(\mbf{k}) \cdot\vec{{\sigma}} + |\mathcal{N}|^2 \sin^2(\kappa_z h) \left[ \vec{d}_h(\mbf{k})-\vec{d}_g(\mbf{k}) \right]\cdot \vec{{\sigma}},
\end{equation}
with $|\mathcal{N}|^2 \sin^2(\kappa_z h)$ representing the amplitude of the $\kappa_z$ state at the HL which has index $h$.
\begin{figure}
\resizebox{0.5\columnwidth}{!}{
\includegraphics[scale=1]{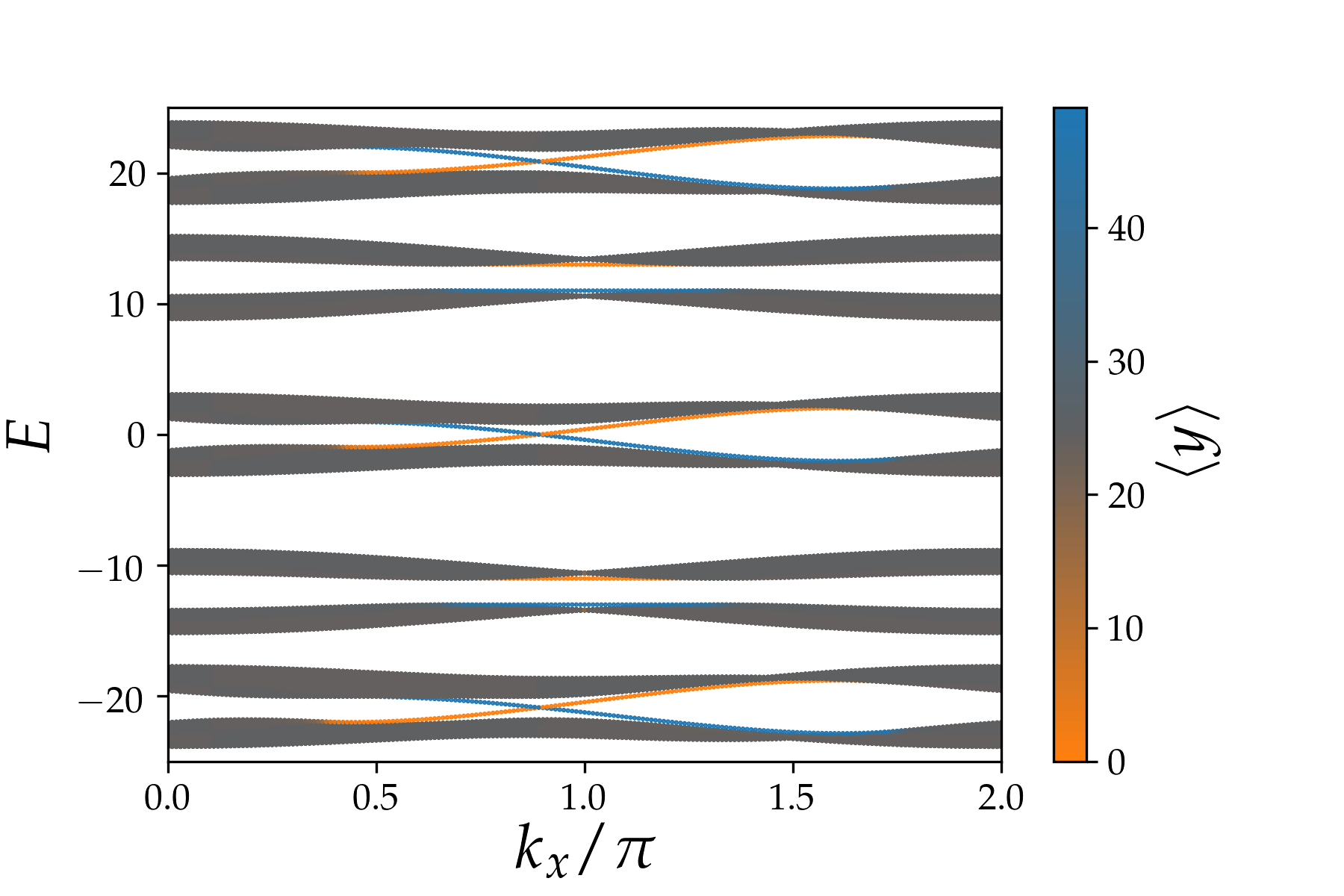}}
\caption{Band structure of a system with $L=5$, $L_1=L_2=2$, $t_1=t_2=m=1$, and $r=12$ with open boundary conditions in $y$ direction. The edges are of zig-zag type. Bands can be split into 5 pairs, each of which is described by slightly different effective Haldane model. The second and fourth pair of bands have no edge state crossing the gap due to $m^\mathrm{eff}/t_2^\mathrm{eff}>3\sqrt{3}$. The remaining pairs have chiral edge states crossing the gap between them. The Fermi energy $E_F=0$ separates two bands of the middle pair resulting in the nontrivial topological properties of a system as a whole.}
\label{fig:app:multi}
\end{figure}
The effective Hamiltonian for given $\kappa_z$ has the form of the standard Haldane model with energy shift $-2r\cos(\kappa_z)$, and new effective parameters $t_1^{\mathrm{eff}}$, $t_2^{\mathrm{eff}}$, $\Phi^{\mathrm{eff}}$, and  $m^{\mathrm{eff}}$. 
For the setup used in the main text we get effective NN hopping amplitude  $t_1^{\mathrm{eff}}=t_1$, staggered potential $m^\mathrm{eff} = \left( 1- |\mathcal{N}|^2 \sin^2(\kappa_z h) \right) m$ and NNN hopping $t_2^\mathrm{eff} = |\mathcal{N}|^2 \sin^2(\kappa_z h) t_2$.
The ratio of these effective parameters $m^\mathrm{eff}$ and $t_2^\mathrm{eff}$ determines the topological properties of a given pair of energy bands.
Note that for $\kappa_z h=n \pi$ with $n\in\mathbb{Z}$ the effective NNN hopping vanishes $t_2^\mathrm{eff}=0$ which means that the state is a dark state.
In Fig.~\ref{fig:app:multi} we present an exaple spectrum for $L=5$, $L_1=L_2=2$, $t_1=t_2=m=1$, and $r=12$ in the cylinder geometry, i.e., periodic boundary conditions in $x$ direction and open ones in $y$ direction.
We observe a clear separation of bands into pairs.
Depending on effective parameters $m^\mathrm{eff}$ and $t_2^\mathrm{eff}$, different for each pair, the bands have Chern numbers being either $C=0$ or $\pm1$.
This is reflected by the edge states shown in blue/orange.
The half-filled system as a whole has a Chern number $C=1$ and a chiral edge state crosses the gap, due to the properties of a middle bands pair, which is separated by the Fermi energy $E_F=0$.

\section{Dark states for arbitrary coupling and number of layers}
\label{app:DS}

The emergence of dark states in multilayer systems described by the Hamiltonian~\eqref{eqn:app:1} can be shown also in the regime of arbitrary coupling strength $r$.
In order to find the condition for their existence we apply a unitary transformation to~\eqref{eqn:app:1} of the form $\mathcal{U}^\dag(\mbf{k}) \H(\mbf{k})\mathcal{U}(\mbf{k})$, where
\begin{equation}\label{eqn:app:6}
\mathcal{U}(\mathbf{k})=\begin{pmatrix}
\ddots & & & & \\
 & \mathbb{U}_g(\mathbf{k}) & \mathbb{0} & \mathbb{0} & \\
 & \mathbb{0} & \mathbb{U}_g(\mathbf{k}) & \mathbb{0} & \\
 & \mathbb{0} & \mathbb{0} & \mathbb{U}_g(\mathbf{k}) & \\
 & & & & \ddots \\
\end{pmatrix},
\end{equation}
and where $\mathbb{U}_g(\mbf{k})$ diagonalizes the graphene Hamiltonian $\vec{d}_g(\mathbf{k}) \cdot \vec{{\sigma}}$.
This transformation leaves coupling blocks with $r$ terms unchanged as they are proportional to the unit matrix, diagonalizes $2\times 2$ blocks which correspond to GLs, and modifies the $2\times 2$ block of the HL in some way, which is irrelevant for the existence of dark states.
Now we consider the GLs above and below the HL. 
That is we look at the first $L_1$ GLs that are coupled by $r$ but not coupled to the HL and the same for for the $L_2$ GLs below the HL.
If these two separate systems share an eigenvalue, then a dark state of the following form exists:
\begin{equation}\label{eqn:app:7}
v_\mathrm{DS} = \left[\alpha \psi_1,\alpha \psi_2, \ldots,\alpha \psi_{2 L_1}, 0, 0,\beta \phi_1, \beta \phi_2, \ldots, \beta \phi_{2L_2}  \right]^T,
\end{equation}
where $\left[\psi_1,\psi_2, \ldots,\psi_{2 L_1}\right]^T$ is an eigenstate of coupled $L_1$ layers above the HL with eigenvalue $\lambda$ and $\left[ \phi_1, \phi_2, \ldots, \phi_{2L_2}  \right]^T$ is an eigenstate of $L_2$ layers below HL, which has the same eigenvalue $\lambda$.
The factors $\alpha$ and $\beta$ are chosen such that the interference is destructive on the HL when applying the Hamiltonian $\mathcal{U}^\dag(\mbf{k}) \H(\mbf{k})\mathcal{U}(\mbf{k})$ to the state in Eq.~\eqref{eqn:app:7}.
In our, $\alpha$ and $\beta$ can always be found, since (i) components $\psi_i$ and $\phi_i$ do not depend on $\mbf{k}$ and (ii) in the eigenbasis of the Hamiltonian $\mathcal{U}^\dag(\mbf{k}) \H(\mbf{k})\mathcal{U}(\mbf{k})$ the states with nonvanishing odd elements $\psi_{2n+1}\neq 0 \neq \phi_{2m+1}$ necessarily have vanishing even elements $\psi_{2n}= 0 = \phi_{2m}$, and vice versa.
Thus it is sufficient to set $\alpha/\beta = -\phi_1/\psi_{2L_1-1}$ or $\alpha/\beta = -\phi_2/\psi_{2L_1}$.

\section{Semimetal}
\label{app:GC}

In this section we show why for the intermediate coupling strengths the gap closes and remains closed for a finite range of $r$ values.
We focus on the three-layer system and HL on top of the two GLs which corresponds to the configuration HGG.
Generalization to configuration GHG and multilayered systems is discussed at the end of this section.
The Hamiltonian is given by Eq.~(1) of the main text.
For $\mbf{k}_d\in\mathcal{D}$, defined in the main text we apply the unitary transformation defined by Eq.~\eqref{eqn:app:6} in the following way:
\begin{equation}\label{eqn:app:9}
\begin{split}
\begin{pmatrix}
 \mathbb{U}^\dag(\mathbf{k}_d) & \mathbb{0} & \mathbb{0} \\
 \mathbb{0} & \mathbb{U}^\dag(\mathbf{k}_d) & \mathbb{0} \\
 \mathbb{0} & \mathbb{0} & \mathbb{U}^\dag(\mathbf{k}_d) \\
\end{pmatrix}
\begin{pmatrix}
\vec{d}_h(\mathbf{k}_d) \cdot \vec{{\sigma}} & r\cdot \mathbb{1} & \mathbb{0} \\
r\cdot \mathbb{1} & \vec{d}_g(\mathbf{k}_d) \cdot \vec{{\sigma}} & r\cdot \mathbb{1} \\
\mathbb{0} & r\cdot \mathbb{1} & \vec{d}_g(\mathbf{k}_d) \cdot \vec{{\sigma}} \\
\end{pmatrix}
\begin{pmatrix}
 \mathbb{U}(\mathbf{k}_d) & \mathbb{0} & \mathbb{0} \\
 \mathbb{0} & \mathbb{U}(\mathbf{k}_d) & \mathbb{0} \\
 \mathbb{0} & \mathbb{0} & \mathbb{U}(\mathbf{k}_d) \\
\end{pmatrix} & \\
=\begin{pmatrix}
\epsilon_+(\mbf{k}_d) & 0 & r & 0 & 0 & 0 \\
0 & \epsilon_-(\mbf{k}_d) & 0 & r & 0 & 0 \\
r & 0 & \epsilon_+(\mbf{k}_d) & 0 & r & 0 \\
0 & r & 0 & \epsilon_-(\mbf{k}_d) & 0 & r \\
0 & 0 & r & 0 & \epsilon_+(\mbf{k}_d) & 0 \\
0 & 0 & 0 & r & 0 & \epsilon_-(\mbf{k}_d)
\end{pmatrix}. & \\
\end{split}
\end{equation}
Here, $\epsilon_\pm(\mbf{k}_d)$ are the eigenvalues of both Haldane and graphene tight-binding model as $\mbf{k}_d\in\mathcal{D}$.
The eigenspace of Eq.~\eqref{eqn:app:9} splits into two orthogonal subspaces related to eigenvalues $\epsilon_+$ and $\epsilon_-$, which satisfy $\epsilon_+ = - \epsilon_-$ for all $\mbf{k}_d$.
The Hamiltonian thus has eigenvalues $E_{\pm,i}(\mbf{k}_d) \in \{ \epsilon_\pm(\mbf{k}_d) +\sqrt{2}r,\ \epsilon_\pm(\mbf{k}_d),\ \epsilon_\pm(\mbf{k}_d) - \sqrt{2}r \}$.
As the subspaces for  $\epsilon_+$ and $\epsilon_-$ are independent, we conclude that at $\sqrt{2}r = \epsilon_+(\mbf{k}_d)$ a degeneracy occurs between states with $\epsilon_+(\mbf{k}_d)-\sqrt{2}r = \epsilon_-(\mbf{k}_d)+\sqrt{2}r = 0$.
We make essential observations for this effect:
\begin{enumerate}
\item along the contour $\mathcal{D}$ of $\mbf{k}_d$ points, the values of $\epsilon_\pm(\mbf{k}_d)$ change smoothly in a periodic manner, which for a given value of $r$ leads to a gap closing at single points rather the entire line of $\mbf{k}_d$'s at the same time,
\item for points outside of $\mathcal{D}$ the above transformation no longer diagonalizes the Haldane model, so a term mixing subspaces $\epsilon_+$ and $\epsilon_-$ keeps the gap open (unless we move far away from the contour, say to $\mathbf{K}$ or $\mathbf{K}'$ point), 
\item there are three pairs of such points which have a low energy dispersion forming Dirac cones,
\item at $r_{<}$ -- the left boundary of a gapless region -- the pairs of these Dirac points emerge at the intersection of the $\mathcal{D}$ contour and the line connecting $\mbf{K}$ and $\mbf{K'}$ points in BZ the. For increasing $r_{<} < r < r_{>}$ these pairs of Dirac points evolve through the BZ along the $\mathcal{D}$ contour in opposite directions, and for $r_{>}$ -- the right boundary of the gapless region -- the Dirac points annihilate in different pairs, as discussed in the main text.
\end{enumerate}

Because the Dirac points at intermediate $r$ are created and annihilated in pairs, it is intuitive to assume that they have opposite vorticity.
Below we provide more rigorous arguments to back up this statement.
First, we consider a fact that the $\mathcal{D}$ contour is symmetric with respect to the axis connecting $\mbf{K}$ and $\mbf{K'}$ points.
We define $\mbf{K}=\frac{2\pi}{3}(1,\frac{1}{\sqrt{3}})$ and $\mbf{K'}=\frac{2\pi}{3}(1,-\frac{1}{\sqrt{3}})$.
Then if we represent one Dirac point as $\mbf{k}_d=(\frac{2\pi}{3}-\tilde{k}_x,k_y)$ we immediately get that $\mbf{k}'_d=(\frac{2\pi}{3}+\tilde{k}_x,k_y)$ will also be a Dirac point.
This is due to the following relation
\begin{equation}\label{eqn:app:10}
\vec{d}_{g/h}(\mbf{k}_d)\cdot\vec{{\sigma}} =
\begin{pmatrix} 1 & 0 \\ 0 & \e^{\rmi\frac{4\pi}{3}}\end{pmatrix}\cdot
\left(\vec{d}_{g/h}(\mbf{k}'_d)\cdot\vec{{\sigma}}\right)^\ast\cdot
\begin{pmatrix} 1 & 0 \\ 0 & \e^{-\rmi\frac{4\pi}{3}}\end{pmatrix}.
\end{equation}
In other words, the $\mbf{k}$-resolved Hamiltonians of the system at $\mbf{k}_d$ and $\mbf{k}'_d$ are related by a complex conjugation combined with a unitary gauge transformation.
Because of the complex conjugation, the two Dirac points have opposite vorticities.
As a result of Dirac points occurring in pairs the gap closing should not affect topological properties of the full system.

In Ref.~\cite{Sorn2018} it is suggested that the above semimetallic properties are lost if one includes next-next-nearest-neighbor (NNNN) hopping $t_3$.
However, contrary to this suggestion, we find evidence that the gap remains closed even in the presence of the NNNN hopping.
This can be seen from the notation of the Hamiltonian using $\vec{d}\cdot\vec{\sigma}$.
In the main text we provide the values of the vector $\vec{d}(\mbf{k})$.
We note that the NNNN hopping couples sites of different sublattices, just as the NN hopping $t_1$.
Therefore, including $t_3$ can only modify elements $x$ and $y$ of the $\vec{d}(\mbf{k})$ vector.
If we include the same NNNN hopping in each layer, the contour given by the condition $\vec{d}_h(\mbf{k})=\vec{d}_g(\mbf{k})$ remains unchanged.
All the above arguments given in this section still hold.
The only difference is observed in the critical values $r_<$ and $r_>$, and in the evolution of Dirac points along the $\mathcal{D}$ contour with increasing $r$.

We close this section with a generalizing remark for multilayer systems.
We notice that the definition of $\mathcal{D}$ is independent of the number of layers and the layer index of the HL.
Therefore, our reasoning seems to also apply to multilayer systems.
However, there are certain differences.
For systems with more than three layers one can expect more than three pairs of Dirac points forming.
On top of that, one can also expect dark states as discussed above.
These can render the full system metallic consisting of two decoupled subsytems, one of which is topological or semi-metallic as discussed in the main text.
Finally, one could consider a system in which different layers differ also with respect to their staggered potential $m$ and NN hopping amplitude $t_1$.
While analytic calculations become much more involved in such case, the results can be obtained numerically.
We found that the presence of the Dirac points in the gapless region is robust to small such perturbations (results obtained for a three-layer system are not shown in here).

\begin{figure}
\resizebox{1.0\columnwidth}{!}{
\includegraphics{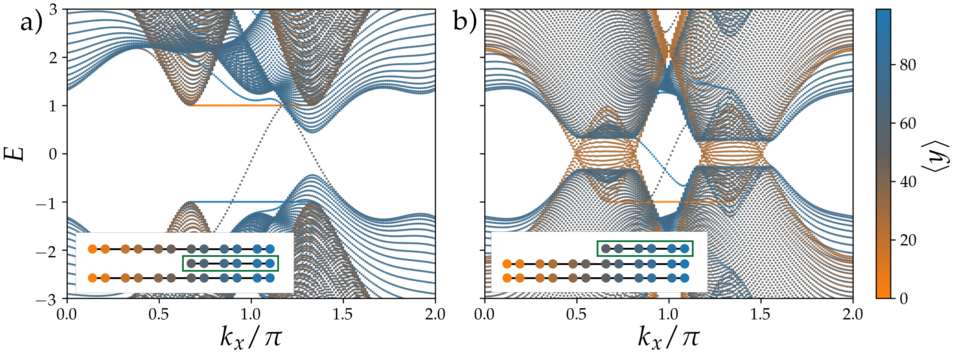}}
\caption{Band structures of the systems with zig-zag edges in the $y$ direction for a) sandwiched HL, b) HL on top of GLs. The boundary at lower values of $y$ is different for HL and GLs, but for higher values of $y$ all layers share the same boundary, as depicted in the insets. Color represents average  position $\langle y \rangle$ of a state along the $y$ axis. We set $r=2$ in both configurations. Other parameters of the system are set as in Fig.~1.}
\label{fig:main:edge}
\end{figure}
\section{Edge state spectra}
We now investigate the bulk-boundary correspondence in the three-layer system.
We choose a system which is infinite in the $x$ direction and has zig-zag edges in the $y$ direction.
Because the edge states of the GLs can couple to edge states of the HL, we employ a similar approach as in Ref.~\cite{Hsieh2016,Cheng2019} in which we take the GLs and the HL of different lengths in $y$ direction which we call \textit{terrace} configuration.
However, in contrast to Ref.~\cite{Cheng2019}, we take the HL to be shorter in order to determine the position of edge state induced in proximity effect on the GLs.
The number of lattice sites in $y$ direction is $N_y^\mathrm{GL}=100$ for the GL and $N_y^\mathrm{H}=50$ for the HL and we set $r=2$ ($t_1$, $t_2$ and $m$ are as in Fig.~1. in the main text), where configuration HGG is topologically nontrivial and configuration GHG is trivial.
The setups are schematically depicted in Fig.~\ref{fig:main:edge}, where we also plot the band structure of the system and in color the average position in $y$ direction $\langle y\rangle$ of each state.

In configuration GHG, Fig.~\ref{fig:main:edge} a), we observe graphene dark state bands which are gapped and decoupled form the rest of the system.
They can be recognized by $\langle y\rangle\approx50$ in grey due to their complete delocalization with respect to the $y$ position.
They also have two flat edge states, characteristic for graphene with zig-zag edges~\cite{Brey2006,Yao2009}.
The dispersion of the dark states is identical to the one of a single GL.
The two bands with distribution similar to the one of the dark states and with $\langle y \rangle \approx 25$ (orange), correspond to the terrace of the GLs, sites with $y<50$ which are not coupled to the HL.
Most importantly, we observe two edge states crossing the gap in the middle of the system $\langle y \rangle\approx 50$.
Contrary to the dark states these states are localized at the edge of the HL.
The edge states reside on all three of the layers, and therefore the bulk topological proximity effect induces an edge state in graphene that, while being localized close to the edge of the intrinsically nontrivial HL, resides in the bulk of graphene.
On the other edge, shared by GL and HL, the gap is not crossed by edge states due to their hybridization.
We conclude that at $r=2$ the system in configuration GHG can either have two counter-propagating edge states or no edge state.
This is in agreement with our bulk investigations, Fig.~1., which predicts the system to be topologically trivial at $r=2$.

In configuration HGG, Fig.~\ref{fig:main:edge} b), we observe that the system becomes metallic.
This is solely due to the geometry of the GLs being longer than the HL.
The metallic properties originate from the terrace parts of two coupled GLs that extend beyond HL.
This can be identified by the orange color corresponding to $\langle y \rangle \approx 25$.
The terrace of the GL has edge states at the $y=0$ edge, characteristic to a zig-zag edge of graphene.
However, they have only one edge state at the onset of the HL, which crosses the gap (of the trilayer part of the system).
The system has also another edge state localized around $y=100$.
In conclusion, the system in configuration GHG has one chiral edge state, as expected from bulk considerations Fig.~1., identified here by the two edge states at the opposite boundaries of the three-layer part of the system, while the metallic terrace can be approximately treated as separate system.
This shows that a topological insulator can exhibit edge states not only at the boundary with a trivial insulator, but also a metallic system (here bilayer graphene).

\bibliographystyle{apsrev4-1}
\bibliography{references}

\end{document}